\documentclass[11pt,epsf]{article}
\usepackage{hyperref}
\usepackage{graphics}
\usepackage{amssymb}
\usepackage{epsf}
\usepackage{epsfig}
\usepackage{rotating}
\usepackage{axodraw}
\usepackage{pstricks}

\newcommand{\veps}{\varepsilon}

\newcommand{\nn}{\nonumber}

\newcommand{\be}{\begin{equation}}
\newcommand{\ee}{\end{equation}}
\newcommand{\bea}{\begin{eqnarray}}
\newcommand{\eea}{\end{eqnarray}}
\newcommand{\ba}{\begin{eqnarray*}}
\newcommand{\ea}{\end{eqnarray*}}

\def\ds#1{#1\kern-1ex\hbox{/}}
\def\dsh{h\kern-1.2ex /}

\def\nn{\nonumber}
\def\beq{\begin{equation}}
\def\eeq{\end{equation}}

\def\beqn{\begin{eqnarray}}
\def\eeqn{\end{eqnarray}}
\def\ba{\begin{eqnarray}}
\def\ea{\end{eqnarray}}

\def\slash#1{#1\hskip-6pt/\hskip6pt}

\newcommand{\beqa}{\begin{eqnarray}}
\newcommand{\eeqa}{\end{eqnarray}}

\newcommand{\si}{\sigma}

\begin{document}

\begin{center}
\vspace{.5cm}
{\bf\large Anomaly Poles as Common Signatures \\ of Chiral and Conformal Anomalies   \\}

\vspace{1.5cm}
{\bf\large $^{(a)}$Roberta Armillis, $^{(a,b)}$Claudio Corian\`{o} and  $^{(a)}$Luigi Delle Rose}
\footnote{roberta.armillis@le.infn.it, claudio.coriano@le.infn.it, luigi.dellerose@le.infn.it}
\vspace{1cm}

{\it  $^{(a)}$Dipartimento di Fisica, Universit\`{a} del Salento \\
and  INFN Sezione di Lecce,  Via Arnesano 73100 Lecce, Italy}\\
\vspace{.5cm}
{\it $^{(b)}$ Department of Physics, University of Crete\\ Heraklion, Crete, Greece}

\begin{abstract}
One feature of the chiral anomaly, analyzed in a perturbative framework, is the appearance of massless poles which account for it. They are identified by a spectral analysis and are usually interpreted as being of an infrared origin. Recent investigations show that their presence is not just confined in the infrared, but that they appear in the effective action under the most general kinematical conditions, even if they decouple in the infrared. Further studies reveal that they are responsible for the non-unitary behaviour of these theories in the ultraviolet (UV) region. We extend this analysis to the case of the conformal anomaly, showing that the effective action describing the interaction of gauge fields with gravity is characterized by anomaly poles that give the entire anomaly and are decoupled in the infrared (IR), in complete analogy with the chiral case. This complements a related analysis by Giannotti and Mottola on the trace anomaly in gravity, in which an anomaly pole has been identified in the corresponding correlator using dispersion theory in the IR. Our extension is based on an exact computation of the off-shell correlation function involving an energy-momentum tensor and two vector currents (the gauge-gauge-graviton vertex) which is responsible for the appearance of the anomaly. 

\end{abstract}
\end{center}
\newpage
\section{Introduction}
In the case of chiral (and anomalous) gauge theories, the corresponding anomalous Ward identities, which are at the core of the quantum formulation of these theories, have a natural and obvious solution, which can be written down quite straightforwardly, in terms of anomaly poles. This takes place even before that any direct computation of the anomaly diagram allows to really identify the presence (or the absence) of such contributions in the explicit expression of an anomalous correlator of the type $AVV$ (A= Axial-Vector, V=Vector) or $AAA$. 

\begin{figure}[t]
\begin{center}
\includegraphics[scale=1.0]{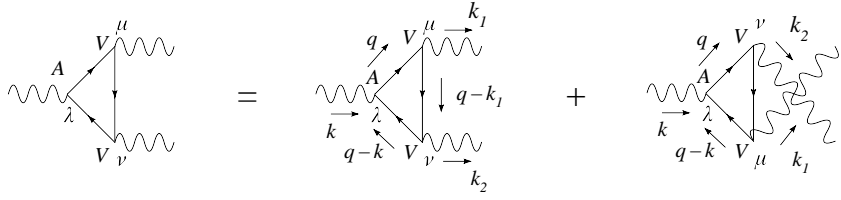}
\caption{\small Triangle diagram and momentum conventions for an AVV correlator.}
\label{AVV}
\end{center}
\end{figure}

To state it simply, the pole appears by solving the anomalous Ward identity for the corresponding amplitude $\Delta^{\lambda\mu\nu} (k_1,k_2) $ (we use momenta as in Fig.~\ref{AVV}  with $k=k_1+k_2$)
\beq
k_\lambda \Delta^{\lambda\mu\nu} (k_1,k_2)= a_n \epsilon^{\mu\nu\alpha\beta}\, k_{1\alpha} \, k_{2\beta}
\eeq
rather trivially, using the longitudinal tensor structure
\beq
\Delta^{\lambda\mu\nu}\equiv w_L= a_n \, \frac{k^{\lambda}}{k^2} \, \epsilon^{\mu\nu\alpha\beta}\, k_{1\alpha} \, k_{2\beta}.
\label{IRpole}
\eeq
In the expression above $a_n = -i / 2 \pi^2$ denotes the anomaly.
The presence of this tensor structure with a $1/k^2$ behaviour is the signature of the anomaly. 
This result holds for an $AVV$ graph, but can be trivially generalized to more general anomaly graphs, such as 
$AAA$ graphs, by adding poles in the invariants of the remaining lines, i.e. $1/k_1^2$ and $1/k_2^2$
\beqa
\Delta^{\lambda \mu \nu}_{AAA}(k,k_1,k_2) &=& \frac{1}{3} \left(
\frac{a_n}{k^2} \, k^\lambda \, \epsilon[\mu, \nu, k_1, k_2]
+ \frac{a_n}{k_1^2} \, k_1^\mu \, \epsilon[\lambda, \nu, k, k_2]
+ \frac{a_n}{k_2^2} \, k_2^\nu \, \epsilon[\lambda, \mu, k, k_1]\right),
\eeqa
imposing an equal distribution of the anomaly on the three axial-vector legs of the graph.

The same Ward identity can be formulated also as a variational equation. The simplest case is that of a theory describing a single anomalous gauge boson $B$ with a Lagrangian
\beq
\mathcal{L}_{B}= \overline{\psi} \left( i \, \slash{\partial} + e \slash{B} \gamma_5\right)\psi - \frac{1}{4} F_B^2,
\label{count0}
\eeq
whose anomalous gauge variation $(\delta B_\mu=\partial_\mu\theta_B)$
\beq
\delta\Gamma_B = \frac{ i \, e^3 \, a_n}{24} \, \int d^4 x \, \theta_B(x) \, F_B\wedge F_B
\label{var1}
\eeq
can be reproduced by the nonlocal action
\beq
{\Gamma}_{pole}= \frac{e^3}{48 \, \pi^2} \, \langle \partial B(x) \square^{-1}(x-y) F_B(y)\wedge F_B(y)
\rangle.
\label{var2}
\eeq
Given a solution of a variational equation, here simplified by Eqs. (\ref{var1}) and (\ref{var2}), it is mandatory to check whether the $1/\square$ (nonlocal) solution is indeed justified by a perturbative computation. 
\begin{figure}[t]
\begin{center}
\includegraphics[scale=0.8]{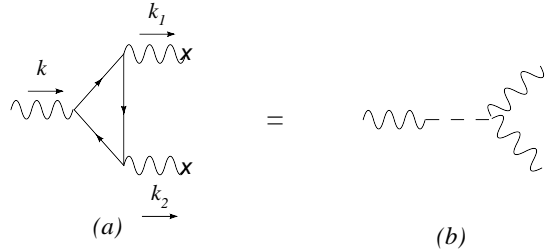}
\caption{\small The amplitude $\Delta^{\lambda\mu\nu}(k_1,k_2)$ shown in $a)$ for the kinematical configuration $k_1^2=k_2^2=0$ reduces to the polar form depicted in $b)$ and given by Eq.~(\ref{IRpole}).}
\label{pole}
\end{center}
\end{figure}
The analysis shows that the kinematical configuration responsible for the appearance of the pole can be depicted as in Fig.~\ref{pole}. In this graph containing the mixing of a spin $1$ with a spin $0$, the anomalous gauge current couples to the two photons via an intermediate massless state which can be interpreted as describing a collinear 
fermion-antifermion pair (a pseudoscalar composite state)  coupled to the two on-shell photons (see also the discussion in \cite{Giannotti:2008cv}). The anomaly graph is characterized, in this limit, by a nonzero spectral density proportional to $\delta(k^2)$ \cite{Dolgov:1971ri}. This kinematical configuration, in which the two photons are on-shell and the fermions are massless, is entirely described by the anomaly pole, which has a clear IR interpretation \cite{Coleman:1982yg}. The IR coupling of the pole present in the correlator is, in this case, rather obvious since the limit
\beq
\lim_{k^2\rightarrow 0} \, k^2 \, \Delta^{\lambda\mu\nu}=k^\lambda \,  a_n \, \epsilon^{\mu\nu\alpha\beta} \, k_{1\alpha} \, k_{2\beta}
\eeq
allows to attribute to this amplitude a non-vanishing residue.

The infrared analysis sketched above is well suited for the identification of anomaly poles which have a rather clear interpretation in this region, but does not allow to identify other similar pole terms which might emerge in far more general kinematical configurations. In \cite{Armillis:2009sm} we have shown that only a complete and explicit computation of the anomalous effective action allows the identification of the extra anomaly poles present in an $AVV$ correlator, that otherwise would escape detection. These have been identified\footnote{A single pole term for an AVV and 3 pole terms for an AAA diagram.}  using a special representation of the anomaly amplitude developed in \cite{Knecht:2003xy,Knecht:2002hr} (that we have called the ``Longitudinal/Transverse" or L/T parameterization), based on the general solution of an anomalous Ward identity. This parameterization takes the form
\beq
 \mathcal \, W ^{\lambda\mu\nu}= \frac{1}{8 \,\pi^2} \left [  \mathcal \, W^{L\, \lambda\mu\nu} -  \mathcal \, W^{T\, \lambda\mu\nu} \right]
\label{long}
\eeq
where the longitudinal component ($W_L$) has a pole contribution ($w_L=-4 i/s$) plus mass corrections 
($\mathcal{F}$) computed in \cite{Armillis:2009sm}
\beq
 \mathcal \, W^{L\, \lambda\mu\nu}= \left(w_L  - \mathcal{F}(m, s,s_1,s_2)\right)  \, k^\lambda \veps[\mu,\nu,k_1,k_2]
\label{brokenW}
\eeq
with
\beq
\mathcal{F}(m, s,s_1,s_2)= \frac{8 \, m^2}{\pi ^2 \, s} \, C_0(s, s_1,s_2,m^2).
\label{due}
\eeq
 The transverse form factors appearing in $W_T$ contribute homogeneously to the anomalous Ward identity. They have been given in the most general case in \cite{Armillis:2009sm}.

Obviously, some doubts concerning the correctness of this parameterization may easily arise, especially if one is accustomed 
to look for anomaly poles using a standard infrared analysis. It is even more so if a pole term of the type shown in Eq.~(\ref{brokenW}) is explicitly present for generic virtualities $s_1$ and $s_2$ of the photons. For this reason and to dissolve any possible doubt, a direct computation shows 
that the L/T representation is, indeed, completely equivalent to the Rosenberg parameterization \cite{Rosenberg:1962pp} of the anomaly graph, even though no poles come to the surface when using this alternative description of the anomaly graph.
In \cite{Armillis:2009sm} one can find an extension of the same parameterization to the massive fermion case, which is indeed given in Eqs. (\ref{brokenW}) and (\ref{due}).  Finally, we have shown that the pole, under general kinematic conditions, is indeed decoupled in the IR. Obviously, 
at this stage, one needs to worry about the precise meaning of this pole, which is explicitly present in some parameterizations, but it is not generated by some special infrared kinematics and as such it does not have a clear IR interpretation.
\section{ Pole-dominated amplitudes} 
\begin{figure}[t]
\begin{center}
\includegraphics[scale=0.8]{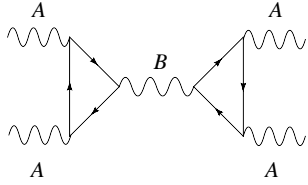}
\caption{\small A ``two-triangles" anomaly amplitude in the $s$-channel which is pole-dominated. In this case we have assumed $A$ to be a non-anomalous gauge boson while $B$ is anomalous.}
\label{BIM}
\end{center}
\end{figure}
A useful device  to investigate the meaning of these new anomaly poles \cite{Armillis:2007tb} is provided by a class of amplitudes \cite{Bouchiat:1972iq} 
which connect initial and final state via anomaly amplitudes, one example of them being shown in Fig.~(\ref{BIM}). These amplitudes are unitarily unbound in the UV \cite{Coriano:2008pg}. This property of theirs can be easily derived by considering the scattering of {\em massless} spin-1 fields coupled via a longitudinal exchange of an anomalous gauge boson. The amplitude in the s-channel is shown in Fig.~\ref{BIM}. In the case of scattering of massless gauge bosons the (IR) pole of Eq.~(\ref{IRpole}) saturates each of the two subamplitudes (i.e. for $m=s_1=s_2=0$). This is an obvious manifestation of the fact that an anomaly 
pole has dangerous effects in the UV due to the broken Ward identity. This behaviour is retained also under general kinematics, for instance in the scattering of massive gauge bosons, when each of the two triangle 
subdiagrams of Fig.~(\ref{BIM}) takes the more general form given by Eqs. (\ref{long}), (\ref{brokenW}). Interestingly enough, if we subtract the pole component contained in $w_L$  \footnote{ We ignore at this point the meaning of this subtraction in the IR. This point is rather delicate and has been discussed in \cite{Armillis:2009sm}  and brings to open ended conclusions concerning the meaning of a ``pole subtraction" scheme.} the quadratic growth of the amplitude disappears \cite{Armillis:2009sm}. Therefore the manifestation of the anomaly and the breaking of unitarity in the UV, in this special kinematical configuration, is necessarily attributed to the $w_L$ component, even if it is decoupled in the IR. After the subtraction, the Ward identity used in the computation of the amplitude remains broken, but it is not anomalous. The apparent breaking of unitarity in the UV is not ameliorated by a more complete analysis of this S-matrix amplitude involving the Higgs sector, since a massless fermion in each of the two anomaly loops would not allow the exchange of a Higgs in the s-channel but the corresponding amplitude would still share the same asymptotic behaviour found for a massive fermion. 

The only possible conclusion extrapolated from this example is that amplitudes which are dominated by anomaly poles in the UV region demonstrate the inconsistency of an anomalous theory, as expected by common lore.
We conclude that unitarity provides a hint on the UV significance of the anomaly poles of the anomaly graphs surfacing in the L/T parameterization, poles which are absent in the usual IR analysis. This does not necessarily exclude a possible (indirect) role played by these contributions in the IR region, nevertheless they do not appear to be artifact generated by the Schouten relation.

The formal solution of the Ward identity \cite{Knecht:2003xy} that takes to the L/T parameterization and to the isolation of an anomaly pole is indeed in agreement with what found in a direct computation. As shown in 
\cite{Armillis:2009sm} one has just to be careful in computing the residue of this parameterization in the IR, where the decoupling of these poles occurs, but it is, for the rest, easy to check.  As $k^2$ is nonzero the separation into longitudinal and transverse contributions is indeed well defined and equivalent to Rosenberg's result \cite{Armillis:2009sm}. These results, as we are going to show, emerge also from the perturbative analysis of the effective action for the conformal anomaly and are likely to correspond to a generic feature of other manifestations of the anomalies in field theory. 
 
\section{The complete anomalous effective action and its expansions in the chiral case}
The point made in \cite{Armillis:2009sm} is that the anomaly is always completely given by $w_L$, under {\em any} kinematical conditions, while the mass corrections (generated, for instance, by spontaneous symmetry breaking) are clearly (and separately) identifiable as extra terms which contribute to the broken anomalous Ward identity satisfied by the correlator. It is important that these two 
sources of breaking of the gauge symmetry (anomalous and spontaneous) be thought of as having both an independent status. For this reason one can provide several organizations of the effective actions of anomalous theories, with similarities that cover both the case of the chiral anomaly and of the conformal anomaly, as we will discuss next. 

The complete effective action, in the chiral case, can be given in several forms. The simplest, valid for any energy range, is the full one
 \beq
 \Gamma^{(3)}=  \Gamma^{(3)}_{pole} + \tilde{\Gamma}^{(3)}
 \eeq
 with the pole part given by (\ref{var2}) and the remainder ( $\tilde{\Gamma}^{(3)}$) given by a complicated nonlocal expression which contributes homogeneously to the Ward identity of the anomaly graph
 \beqa
  \tilde{\Gamma}^{(3)}&=&  -  \frac{e^3}{48 \pi^2} \int d^4 x \, d^4 y \, d^4 z \,  
  \partial \cdot B(z) F_B (x) \wedge F _B (y )
  \int \frac{d^4  k_1 \, d^4 k_2}{(2 \pi)^8} \, 
 e^{-i k_1 \cdot (x-z) - i k_2 \cdot (y-z)} \mathcal F (k, k_1, k_2, m)  \nn \\
&& -\frac{e^3}{48 \pi^2} \int d^4 x \, d^4 y \, d^4 z \, B_\lambda (z) \, B_\mu (x) \, B_\nu (y) 
  \int  \frac{d^4  k_1 \, d^4 k_2}{(2 \pi)^8} \, e^{-i k_1 \cdot (x-z) - i k_2 \cdot (y-z)} \, W_T ^{\lambda \mu \nu} (k, k_1,k_2,m). \nn \\
\label{gammafull}
 \eeqa
 The expressions of these form factors can be found in \cite{Armillis:2009sm}. This (rather formal) expression is an exact result, but becomes more manageable if expanded in the fermion mass (in $1/m$ or in $m$) 
 (see for example \cite{Bastianelli:2007jv, Bastianelli:2004zp}).

For instance, let's consider the $1/m$ case. One of the shortcomings of this expansion, as we are going to argue next, is that it does not do full justice 
of the presence of massless degrees of freedom in the theory (anomaly poles do not appear explicitly in this expansion) which, as discussed in \cite{Giannotti:2008cv} might instead be of physical significance since they are not connected to any scale. 

A second expansion of the effective action 
Eq.~(\ref{gammafull}) can be given for a small mass $m$ (in $m^2/s$). In this formulation the action is organized in the form of a pole contribution plus $O(m^2/s)$ corrections. In this case it is not suitable to describe the heavy fermion limit, but the massless pseudoscalar degrees of freedom introduced by the anomaly in the effective theory can be clearly identified from it. As discussed in \cite{Armillis:2009sm, Coriano:2008pg, Armillis:2008bg}   these are: one axion and one ghost. This expansion gives ($s<0$) 
\beq
w_L =  - \frac{4 i}{s}  - \frac{ 4 \, i \, m^2}{s^2} \log \left( - \frac{s}{m^2}\right) + O (m^3)
\label{smooth}
\eeq
which has a smooth massless limit. It seems to us that this form of the effective action is the most suitable for the study of the UV behaviour of an anomalous theory, in the search of a possible UV completion. Notice that the massless limit of this action reflects (correctly) the pole-dominance present in  the theory in the UV region 
of $s\to \infty$, since the mass corrections are suppressed by $m^2/s^2$.

\section{The conformal anomaly case} 
While this intriguing pattern of pole dominance in the UV and of decoupling in  the IR (for massive or off-shell correlators) is uncovered only after a  complete perturbative analysis of the general 
anomaly graph, it is not just a property of the chiral case. As we are going to show, 
a similar behaviour is typical of the conformal anomaly. We summarize the results of our analysis, details will be given elsewhere \cite{newpaper}. 

In a recent work \cite{Giannotti:2008cv} Mottola and Giannotti have shown that the diagrams responsible for the generation of the conformal anomaly contain an anomaly pole. In their analysis they classify the form factors of the correlator which is responsible for the conformal anomaly graph, which is the photon-photon-graviton vertex, or $TJJ$ correlator, involving the vector current (J) and the energy-momentum tensor (T). The authors use a Ward identity that enforces conservation of the energy-momentum tensor to fix the correlator, which can also be fixed by imposing the general form of the trace anomaly in the massive fermion case. Their analysis shows conclusively that anomaly poles can be extracted in the IR using dispersion theory, similarly to the chiral case. This point had also been noticed in \cite{Horejsi:1997yn} in the study of the 
Ward identity of the correlators describing the trace anomaly at zero momentum transfer.

The identification of these contributions is relevant for establishing the correct expression of the gauge related terms in the gravitational effective action. The spectral analysis of \cite{Giannotti:2008cv} proves that variational solutions of the trace anomaly equation that will be given below in Eq.~(\ref{riegert}), indeed, correctly account at least for some of the contributions to the effective action of these theories. Mass-dependent corrections and other traceless terms which are not part of the anomaly, of course, are not identified by this solution.

We recall that the gravitational trace anomaly in 4 spacetime dimensions generated by quantum effects in a classical gravitational and electromagnetic background is given by the expression
\beq
T_\mu^\mu= -\frac{1}{8} \left( 2 b \,C^2 + 2 b'( E - \frac{2}{3}\square R) + 2 c\, F^2\right),
\eeq
where the $b$ and $b'$ and $c$ are parameters. $C^2$ denotes the Weyl tensor squared and $E$ is the Euler density given by
\beqa
C^2 &=& C_{\lambda\mu\nu\rho}C^{\lambda\mu\nu\rho} = R_{\lambda\mu\nu\rho}R^{\lambda\mu\nu\rho}
-2 R_{\mu\nu}R^{\mu\nu}  + \frac{R^2}{3}, \\
E &=& ^*\hskip-.1cm R_{\lambda\mu\nu\rho}\,^*\hskip-.1cm R^{\lambda\mu\nu\rho} =
R_{\lambda\mu\nu\rho}R^{\lambda\mu\nu\rho} - 4R_{\mu\nu}R^{\mu\nu}+ R^2.
\eeqa
 For a single fermion in the theory we have that $b = 1/320\pi^2$, and $b' = - 11/5760\pi^2$
and $c= -e^2/24\pi^2$.

The effective action, in this approach, is identified by solving the variational equation by inspection, similarly to what we have discussed in the previous section in the case of the chiral anomaly. In this case the equation takes the form
\beq
-\frac{2}{\sqrt{g}}g_{\mu\nu} \frac{\delta \Gamma}{\delta g_{\mu\nu}}=T_\mu^\mu.
\label{traceq}
\eeq
The solution of this variational equation is well known and is given by the nonlocal expression \cite{Riegert:1984kt}
\beqa
&& \hspace{-.6cm}S_{anom}[g,A] = \label{Tnonl} \nn \\
&&\frac {1}{8}\int d^4x\sqrt{-g}\int d^4x'\sqrt{-g'} \left(E - \frac{2}{3} \square R\right)_x
 G_4(x,x')\left[ 2b\,C^2
 + b' \left(E - \frac{2}{3} \square R\right) + 2c\, F_{\mu\nu}F^{\mu\nu}\right]_{x'}. \nonumber \\
\label{riegert}
\eeqa
 The notation $G_4(x,x')$ denotes the Green's function of the
differential operator defined by
\beq
\Delta_4 \equiv  \nabla_\mu\left(\nabla^\mu\nabla^\nu + 2 R^{\mu\nu} - \frac{2}{3} R g^{\mu\nu}\right)
\nabla_\nu = \square^2 + 2 R^{\mu\nu}\nabla_\mu\nabla_\nu +\frac{1}{3} (\nabla^\mu R)
\nabla_\mu - \frac{2}{3} R \square\,
\label{operator4}
\eeq
and requires some boundary conditions to be specified.
 The nonlocal action shows the presence of  a massless pole in the linearized limit  \cite{Giannotti:2008cv}
\beq
S_{anom}[g,A] = -\frac{c}{6}\int d^4x\sqrt{-g}\int d^4x'\sqrt{-g'}\, R^{(1)}_x
\, \square^{-1}_{x,x'}\, [F_{\alpha\beta}F^{\alpha\beta}]_{x'}\,,
\label{simplifies}
 \eeq
 valid for a weak gravitational field ($g_{\mu\nu}= \eta_{\mu\nu} + \kappa h_{\mu\nu}$, $\kappa^2=16 \pi G$). In this case 
 \beq
 R^{(1)}_{\mu\nu}\equiv \partial^x_\mu\partial^x_\nu h^{\mu\nu} - \square h, \qquad h=\eta_{\mu\nu} h^{\mu\nu}.
 \eeq
Eq.~(\ref{simplifies}) can be reproduced by a perturbative analysis.
\section{The $TJJ$ correlator}
\begin{figure}[t]
\begin{center}
\includegraphics[scale=0.9]{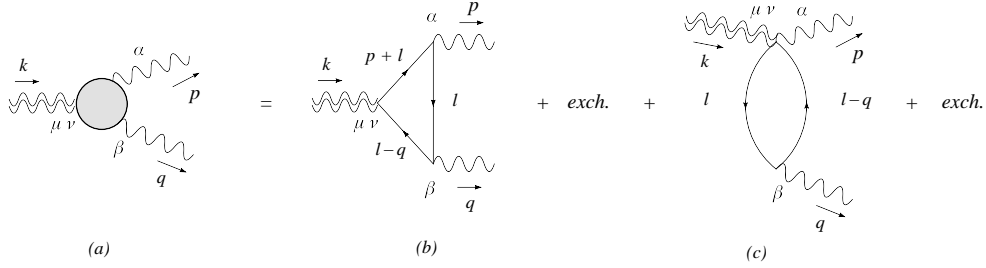}
\caption{\small The complete one-loop vertex  $\Gamma^{\mu \nu \alpha\beta}$ in $a)$ obtained as the sum of two 1PI contributions in $b)$ and $c)$ and of their Bose symmetric diagrams. }
\label{vertex}
\end{center}
\end{figure}
To clarify this point we consider the linearized expression of the gauge contribution to the gravitational effective action which is given by
\beq
S_{TJJ} = \int d^4 x d^4 y d^4 z \, \Gamma^{\mu \nu \alpha\beta}(x,y,z) A_{\alpha}(x) \, A_{\beta}(y) \, h_{\mu \nu}(z), 
\label{STJJ}
\eeq
with  $\Gamma^{\mu \nu \alpha\beta}(x,y,z)$ being the expression of the correlator of two gauge currents with an extra insertion of the energy-momentum tensor (see Fig.~\ref{vertex}) at nonzero momentum transfer. We discuss the QED case. We recall that the coupling to gravity of QED, in the weak field limit, is described by the total energy-momentum tensor via an interaction of the form $h_{\mu\nu} T^{\mu\nu}$ where 
\beq
T^{\mu\nu}\equiv T_{Dirac}^{ \mu\nu} +T_{int}^{ \mu\nu} + T_{e.m.}^{ \mu\nu}.
\eeq
In this case, specifically, one has
\bea
T^{\mu\nu}_{Dirac} &=& -i \bar\psi \gamma^{(\mu}\!\!
\stackrel{\leftrightarrow}{\partial}\!^{\nu)}\psi + g^{\mu\nu}
(i \bar\psi \gamma^{\lambda}\!\!\stackrel{\leftrightarrow}{\partial}\!\!_{\lambda}\psi
- m\bar\psi\psi), \\
\label{tfermionic}
T^{\mu\nu}_{int} &= &-\, e J^{(\mu}A^{\nu)} + e g^{\mu\nu}J^{\lambda}A_{\lambda} , \\
T^{\mu\nu}_{ e.m.} &=& F^{\mu\lambda}F^{\nu}_{\ \ \lambda} - \frac{1}{4} g^{\mu\nu}
F^{\lambda\rho}F_{\lambda\rho},
\label{tphoton}
\eea
where the current is given by
\beq
J^{\mu}(x) = \bar\psi (x) \gamma^{\mu} \psi (x)\,.\\
\label{vectorcurrent}
\eeq
We have introduced some standard notation for the symmetrization of the tensor indices and left-right derivatives
$H^{(\mu\nu)} \equiv (H^{\mu\nu} + H^{\nu\mu})/2$ and
$\stackrel{\leftrightarrow}{\partial}\!\!_{\mu} \equiv
(\stackrel{\rightarrow}{\partial}\!\!_{\mu} - \stackrel{\leftarrow}{\partial}\!\!_{\mu})/2$.

The amplitude present in Eq.~(\ref{STJJ}) can be expanded in a specific base given by
\bea
\Gamma^{\mu\nu\alpha\beta}(p,q) =  \, \sum_{i=1}^{13} F_i (s; s_1, s_2,m^2)\ t_i^{\mu\nu\alpha\beta}(p,q)\,,
\label{Gamt}
\eea
where the $13$ invariant amplitudes $F_i$ are functions of the kinematical invariants $s=k^2=(p+q)^2$, \mbox{$s_1=p^2$}, $s_2=q^2$ and of the internal mass $m$. In \cite{Giannotti:2008cv} the authors use the
Feynman parameterization and momentum shifts in order to identify the expressions of these amplitudes in terms of parametric integrals. This was also the approach followed by Rosenberg in his original identification  of the 6 invariant amplitudes of the AVV anomaly diagram\footnote{The explicit expression of the Rosenberg's integrals have been given  in \cite{Armillis:2009sm}}. The list of amplitudes  $F_i$  can be found in \cite{Giannotti:2008cv}  together with the expressions of the tensors $t_i^{\mu\nu\alpha\beta}(p,q)$. The number of these form factors reduces from $13$ to $3$ in the case of on-shell photons, as shown long ago by Berends and Gastmans  \cite{Berends:1975ah}.  For our purposes, the only amplitudes contributing to the trace anomaly in the massive fermion case come from the tensors $t_1^{\mu\nu\alpha\beta}$ and $t_2^{\mu\nu\alpha\beta}$. They are given by 
\beqa	
t_1^{\mu\nu\alpha\beta}&=&\left(k^2 g^{\mu\nu} - k^{\mu } k^{\nu}\right) u^{\alpha\beta}(p,q), \\
t_2^{\mu\nu\alpha\beta}&=&\left(k^2 g^{\mu\nu} - k^{\mu } k^{\nu}\right) w^{\alpha\beta}(p,q),
\eeqa
where
\bea
&&u^{\alpha\beta}(p,q) \equiv (p\cdot q) g^{\alpha\beta} - q^{\alpha}p^{\beta}\,,\nonumber \\
&&w^{\alpha\beta}(p,q) \equiv p^2 q^2 g^{\alpha\beta} + (p\cdot q) p^{\alpha}q^{\beta}
- q^2 p^{\alpha}p^{\beta} - p^2 q^{\alpha}q^{\beta}.\,
\label{uwdef}
\eea
The identification of an anomaly pole which is not of clear IR origin requires an explicit computation of the effective action, which is rather involved in perturbation theory, but we omit details and just summarize the relevant results. We perform a computation with the kinematical constraint $s_1=s_2=0$ (i.e. two on-shell photons) and a massive fermion. We obtain
\beq
{F_1 (s;\,0,\,0,\,m^2)} = 
F_{1\, pole} \,  + \, \frac{e^2 \,   m^2}{3 \, \pi ^2 \, s^2} \, - \frac{e^2 \, m^2}{3 \, \pi^2 \, s}  \, \mathcal C_0 (s, 0, 0, m^2) \bigg[\frac{1}{2}-\frac{2 \,m^2}{ s}\bigg],  \nn \\
\label{oom}
\ee
where 
\beq F_{1\, pole}=- \frac{e^2 }{18 \, \pi^2 s} 
\eeq
 and the scalar three-point function $ \mathcal C_0 (s, 0,0,m^2) $ is given by 

\beq
\mathcal C_0 (s, 0,0,m^2) = \frac{1}{2 s} \, \log^2 \frac{a_3+1}{a_3-1}, 
\eeq
with $a_3 = \sqrt {1-4m^2/s}$.
The form factor $F_2$, which in general gives a nonzero contribution to the trace anomaly in the presence of mass terms, is multiplied by a tensor structure ($t_2$) which vanishes when the two photons are on-shell. It is quite straightforward to figure out that the pole term ($F_{1 pole}$) given above corresponds to a contribution to the gravitational effective action of the form (\ref{simplifies}), with a linearized scalar curvature. 
Therefore, similarly to the case of the chiral anomaly, also in this case the anomaly is entirely given by $F_{1 pole}$, even in a configuration which is not obtained from a dispersive approach. The presence of mass corrections in 
(\ref{oom}) is not a source of confusion, since there is a clear separation between anomaly and non-conformal breakings of the conformal symmetry.
\subsection{$F_1$ in the most generic case}
A similar result is found in the most general case. After defining  $\gamma \equiv s -s_1 - s_2$  and 
$\si \equiv s^2 - 2 (s_1+s_2)\, s + (s_1-s_2)^2$ we obtain
\bea
F_1 (s;\,s_1,\,s_2,\,m^2) &=& F_{1\, pole}
  + \frac{ e^2 \,  \gamma  \,  m^2}{3 \pi^2 \, s \sigma } 
  + \frac{ e^2 m^2 \, s_2}{3 \pi^2 \,s \, \sigma^2 }  \mathcal D_2(s,s_2,m^2)\,  
  \left[s^2+4 s_1 s-2 s_2 s-5 s_1^2+s_2^2+4 s_1   s_2\right]
    \nn \\
&&  
 -  \, \frac{ e^2 \, m^2 \, s_1}{3 \, \pi^2 \,s \, \sigma^2 } \, \mathcal D_1 (s,s_1,m^2)\,
   \left[-\left(s-s_1\right){}^2+5 s_2^2-4 \left(s+s_1\right)
   s_2\right]   \nn \\
  && \hspace{-3cm}  
  - \, \frac{ e^2\, m^2 \, \gamma}{6 \, \pi^2 \, s \, \sigma^2 } \, \mathcal C_0 (s,s_1,s_2,m^2)\, \left[\left( s-s_1\right){}^3 - s_2^3 + \left(3 s+s_1\right)
   s_2^2 + \left(-3 s^2-10 s_1 s+s_1^2\right) s_2  - 4 m^2 \sigma \right], \nonumber \\
  \label{Fone}
 \eeqa
 
 where
\beq
 \mathcal D_i \equiv \mathcal D_i (s, s_i,  m^2) = \left[ a_i \log\frac{a_i +1}{a_i - 1}
- a_3 \log \frac{a_3 +1}{a_3 - 1}  \right],  \qquad a_i = \sqrt{ 1 - \frac{4m^2}{s_i}}.
\label{D_i}
\eeq
It is quite obvious from the most general expression of $F_1$ that the massless pole, which accounts for the entire trace anomaly, is indeed part of the spectrum.
 The pole decouples in the infrared, as one can show after a detailed study of the entire correlation function  $\Gamma^{\mu\nu\alpha\beta}(p,q)$.
 
 There is something to learn from perturbation theory: anomaly poles are not just associated to the collinear fermion-antifermion limit of the amplitude, but are also present in other, completely different kinematical domains where the collinear kinematics is not allowed and are not detected using a dispersive approach. They are present in the off-shell effective action as they are in the on-shell ones. Proving their decoupling in the IR requires a complete analysis of the anomalous contributions to the effective action, along the lines of \cite{Armillis:2009sm}. 

\section{Lessons from the $1/m$ expansions}
One obvious question to ask is if the nonlocal structure of the poles, which accounts for the anomaly also in the case of the conformal anomaly, is not clearly visible in a given operatorial expansion in terms of higher-dimensional operators.  This is indeed the case, for instance, if we decide to expand in $1/m$  the form factor 
$F_1$. We obtain
\beq
F_{1} (s, 0, 0, m^2) = 
\frac{7 \, e^2 }{135 \cdot 16 \, \pi ^2} \frac{1}{m^2} 
 + \frac{ e^2  \,  s}{189 \cdot 16 \pi ^2}  \frac{1}{m^4} +O\left(\frac{1}{m^6}\right), 
\label{1om}
\eeq
with no signature of the presence of non-decoupling contributions in the UV, which are scaleless and described entirely by the anomaly pole. Of course $1/m$ expansions are legitimate, but there is no apparent sign left in \ref{1om} of the presence of a massless contribution to the conformal anomaly, due to the universal appearance of a mass term.  Another important observation is that the contributions to the trace of the energy-momentum tensor, which is relevant in the cosmological context \cite{Starobinsky:1980te, Shapiro:2006sy},  are all dominated by the pole term at high energy, since mass corrections contained in $F_1$ are clearly suppressed as $m^2/s$. Obviously, Eq.~(\ref{1om}) differs systematically from the result obtained from the small $m$ expansion, where the nonlocality of the effective action and the presence of a massless pseudoscalar exchange, as a result of the conformal anomaly, 
is instead quite evident. We obtain in this second case
\bea
F_1(s,0,0,m^2) = F_{1 \, pole}   
+ \frac{ e^2 \, m^2}{12 \, \pi^2 \, s^2} 
\left[ 4 -  \log^2 \frac{m^2}{s} - 2 i \pi \log \frac{m^2}{s} + \pi^2 	\right] + O \left( \frac{1}{s^3} \right)
\eea
where the anomalous form factor shows a massless pole beside some additional mass corrections.
This is an expansion, as in the case of the chiral anomaly, which is also useful in the UV limit. It appears to  be closer to the complete result even for a large fermion mass, since it keeps the two sources of breaking of the conformal symmetry separated. In this respect it would probably be of interest to see whether the  effects of superluminality \cite{Shore:2003zc} in a weak (external) gravitational field, found in the $1/m$ expansion of the effective action of \cite{Drummond:1979pp}, have anything to share with the presence of massless poles in the effective description.

\section{Conclusions} 
The presence of anomaly poles in perturbation theory appears to be an essential property of anomalous theories, even in the most general kinematical configurations of the anomalous correlators. We have reviewed previous work on the study of the anomaly poles of anomalous gauge currents, with the intent to show the similarities between chiral and conformal anomalies. Our explicit computation, in the case of the trace anomaly, shows that pole singularities appear also in non-collinear configurations of the corresponding anomaly graphs. As we have stressed, these poles are not identified by a spectral analysis but their existence should not be matter of controversy. Historically,  the signature of the anomaly has been attributed to a pole in the anomalous correlator only in the IR region. Our conclusions, contained in a previous work, were
that anomaly poles are instead generic, and not artifacts of a given parameterization or due to the presence of the Schouten relations. Here, building on more recent studies of the conformal anomaly in perturbation theory, we have shown that the perturbative signature of a conformal anomaly is, again, an anomaly pole and that the correlator responsible for the conformal anomaly has properties which are typical of the gauge anomaly.
The pole, also in this case, can be coupled or decoupled in the IR, and raises significant questions concerning the significance and the implications of massless scalar degrees of freedom in gravity, recently addressed in \cite{Giannotti:2008cv}.  
In the case of anomalous chiral gauge theories similar issues  \cite{Coriano:2008pg} \cite{Armillis:2009sm} have been raised concerning the significance of massless pseudoscalar degrees of freedom (gauged axions) and their correct interpretation in a simple field theory language.

  \centerline{\bf Acknowledgements}
We thank M. Guzzi for discussions. This work is supported in part  by the European Union through the Marie Curie Research and Training Network ``Universenet'' (MRTN-CT-2006-035863).  
C.C. thanks the Theory Group at Crete for hospitality and partial financial support by the EU grant INTERREG IIIA (Greece-Cyprus) and FP7-REGPOT-2008-1-CreteHEPCosmo-228644.

\end{document}